\documentclass{jaa}

\usepackage{graphicx}
\usepackage{color}

\begin{document}


\title{Study of advective accretion flow properties around rotating black holes - Application  to GRO J1655-40}



\author{Ramiz Aktar\textsuperscript{1,*}, Santabrata Das\textsuperscript{1}, Anuj Nandi\textsuperscript{2} \and H. Sreehari\textsuperscript{2,3}}
\affilOne{\textsuperscript{1}Indian Institute of Technology Guwahati, Guwahati, Assam, 781039, India.\\}
\affilTwo{\textsuperscript{2}Space Astronomy Group, ISITE Campus, ISRO Satellite Centre, Outer Ring Road, Marathahalli, Bangalore, 560037, India.\\}
\affilThree{\textsuperscript{3}Indian Institute of Science, Bangalore, 560012, India.}


\twocolumn[{

\maketitle

\corres{ramiz@iitg.ernet.in (RA)}

\msinfo{1 January 2015}{1 January 2015}{1 January 2015}

\begin{abstract}
We examine the properties of the viscous dissipative accretion flow around rotating black 
holes in presence of mass loss. Considering thin disc approximation, we self-consistently 
calculate the inflow-outflow solutions and observe that the mass outflow rates decreases 
with the increase of viscosity parameter ($\alpha$). 
Further, we carry out the model calculation of Quasi-periodic Oscillation frequency 
($\nu_{\rm QPO}$) that is frequently observed in black hole sources and observe that
$\nu^{\rm max}_{\rm QPO}$ increases with the increase of black hole spin ($a_k$). Then, we employ
our model in order to explain the High Frequency Quasi-periodic Oscillations (HFQPOs) 
observed in black hole source GRO J1655-40. While doing this, we attempt to constrain the 
range of $a_k$ based on observed HFQPOs ($\sim$ 300 Hz and $\sim$ 450 Hz) for the 
black hole source GRO J1655-40.
\end{abstract}


}]


\doinum{12.3456/s78910-011-012-3}
\artcitid{\#\#\#\#}
\volnum{123}
\year{2017}
\pgrange{23--25}
\setcounter{page}{23}
\lp{25}

\section{Introduction}
Jets and outflows are ubiquitous in accreting black hole systems such as stellar mass 
black holes in X-ray binaries (XRBs), intermediate mass black holes (IMBHs) and super
massive black holes (SMBHs), respectively. By now, observations as well as numerical
simulations confirm the existence of jets and outflows from accreting black hole systems
(Mirabel et al. 1992; Mirabel \& Rodriguez 1994; Mirabel \& Rodriguez 1998; Fender et al. 2009; Miller et al. 2012b; 
Yuan et al. 2012, 2015; Das et al. 2014; Okuda \& Das 2015; 
Mezcua et al. 2013, 2015; Fender \& Munoz-Darias 2016). As the black holes do not have any 
hard surfaces or boundaries,  the outflows or jets must be emerged out from the 
accretion disc itself. Meanwhile, several observations confirm the existence of disk-jet 
connections in black hole systems (Feroci et al. 1999; Vadawale et al. 2001; Nandi et al. 
2001; Gallo et al. 2003; Miller et al. 2012a; Miller-Jones et al. 2012; Corbel et al. 
2013; Radhika \& Nandi 2014; Radhika et al. 2016a, Radhika et al. 2016b).

It is well know that in an advective accretion disc around black hole, accreting matter
must be transonic in nature in order to satisfy the inner boundary condition imposed 
by the black hole horizon. Moreover, during the course of accretion, rotating  matter
experiences centrifugal repulsion in the vicinity of the black hole. When repulsion 
become comparable against the attractive gravity force, accreting matter slows down
and starts to pile up that eventually triggers the discontinuous transition of flow 
variables in the form of shock wave. As a result, a virtual barrier is formed
around the black hole that effectively behaves like the boundary layer of the black hole
and we call it as post-shock corona (hereafter PSC). Interestingly, the formation of PSC 
in an accretion flow is thermodynamically preferred as it possesses high entropy 
content (Becker \& Kazanas 2001). Following this idea, the existence of 
shock wave around the black holes and its physical implications have been studied 
by several group of authors (Fukue 1987; Chakrabarti 1989; Chakrabarti 
1996b; Lu et al. 1999; Becker \& Kazanas 2001; Das et al. 2001a; Chakrabarti \&
Das 2004; Fukumura \& Tsuruta 2004;  Mondal \& Chakrabarti 2006; Das 2007; 
Das \& Chakrabarti 2008; Das et al. 2009; Das et al. 2010; Aktar et al. 2015; 
Sarkar \& Das 2016; Aktar et al. 2017). When the flow is being advected towards 
the black hole, a part of the infalling matter after being intercepted at PSC is 
deflected along the black hole rotation axis to form bipolar outflows (Das et al., 2014, 
and reference therein).

In a theoretical attempt, Penrose (1969) and Blandford \& Znajek (1977) proposed 
that the powerful jets could be originated due to the spin of the black hole.  
But, conflicting claims are also reported in the literature afterwards. 
Steiner et al. (2013) and 
McClintock et al. (2014) observed a significant positive correlation between the jet 
power and spin of the black hole. On the other hand, Russel et al. (2013) and Fender 
\& Gallo (2014) claimed that there are no such correlation exist based on their 
observational studies. Recently, Aktar et al. (2015) calculated maximum outflow rates 
as a function of spin of the black hole and observed no such significant positive 
correlation between spin and maximum outflow rates based on their accretion-ejection 
model. Essentially, in the context of black hole system,
the spin-jet correlation is not conclusive yet.

With the above findings, in this work, we consider an accretion-ejection model
around a rotating black hole and investigate the effect of viscous dissipation on the 
generation of mass outflow rates. Further, we employ our model calculation to obtain
$\nu_{\rm QPO}$. We identify the 
maximum value of QPO frequency ($\nu^{\rm max}_{\rm QPO}$) and find that 
$\nu^{\rm max}_{\rm QPO}$ increases with the increase of black hole spin ($a_k$). 
Next, we put an effort to explain the origin of HFQPOs for the 
source GRO J1655-40. For the purpose of representation, we consider two HFQPOs, 
namely $\sim 300$ Hz and $\sim 450$ Hz that was observed in GRO J1655-40 and using our model 
calculation, we constrain the range of $a_k$ and $\alpha$ that successfully reproduces
the above two HFQPOs for this black hole source.

In Section 2, we present the governing equations for accretion and outflow. In section 3, we 
discuss the solution methodology. In section 4, we discuss the obtained results. 
In section 5, we make use of the observed HFQPOs in GRO J1655-40 source 
to constrain the spin of the black hole. Finally, in section 6, we present
the conclusion of our work.

\section{Governing equations for accretion and outflow}
In this paper, we consider a disc-jet system around a rotating black hole. The 
accretion disc lies along the equatorial plane and the jet geometry is considered in 
the off-equatorial plane about the axis of rotation of the black hole (Molteni et al. 
1996a, Aktar et al. 2015, Aktar et al. 2017). Further, we consider a steady, 
geometrically thin, axisymmetric, viscous accretion flow around 
rotating black hole. We adopt pseudo-Kerr potential of Chakrabarti \& Mondal 
(2006) to mimic the space-time geometry around black hole. Here, we assume the 
unit system as $G=M_{BH}=c=1$ to represent the flow variables, where, $G$, 
$M_{BH}$ and $c$ are the gravitational constant, mass of the black hole and speed 
of light, respectively.

\subsection{Equations for Accretion}
The equations for accretion are given by,\\
(i) The radial momentum equation:
$$
v\frac{dv}{dx} + \frac{1}{\rho}\frac{dP}{dx} + \frac{d\Phi_{eff}}{dx} =0 ,
\eqno(1)
$$ 
where, $v$, $P$, $\rho$ and $x$ are the speed of matter, gas pressure, density and 
radial distance of the accretion flow, respectively. Here, $\Phi_{eff}$ is the effective 
potential around black hole. 
This potential mimic the Kerr geometry quite satisfactorily within the range of $-1\leq 
a_k\leq0.8$ (Chakrabarti \& Mondal 2006).

(ii) The mass conservation equation:
$$
\dot{M}=4 \pi \rho v x h ,
\eqno(2)
$$
where, ${\dot M}$ denotes the mass accretion rate which is constant throughout the 
flow except at the region of mass loss. Here, $4\pi$ is the geometric constant. We 
calculate the half-thickness of the disc $h(x)$ considering the
hydrostatic equilibrium in the vertical direction and is given by,
$$
h(x)=a\sqrt{\frac{x}{\gamma \Phi_{r}^{\prime}}},
\eqno(3)
$$
where, $a$ is the adiabatic sound speed defined as $a=\sqrt{\frac{\gamma P}
{\rho}}$ and $\gamma$ is the adiabatic index. Here, $\Phi_{r}^{\prime}=\left( 
\frac{\partial \Phi_{eff}}{\partial r}\right)_{z<<x}$, and $z$ is
the vertical height in the cylindrical coordinate system where $r = \sqrt{x^2 + z^2}$.

(iii) The angular momentum distribution equation:
$$
v\frac{d\lambda}{dx} + \frac{1}{\Sigma x}\frac{d}{dx}(x^2 W_{x\phi}) =0.
\eqno(4)
$$
where, $W_{x\phi}$ is the viscous stress which is dominated by the $x\phi$ 
component. The viscous stress $W_{x\phi}$ is modeled according to Chakrabarti 
(1996a) as,
$$
W_{x\phi} = -\alpha (W + \Sigma v^2),
\eqno(5)
$$
where,  $W = 2I_{n+1}Ph(x)$ and $\Sigma = 2 I_n \rho h(x)$ are the vertically 
integrated pressure and density (Matsumoto et al. 1984). Here, $I_n$ and $I_{n+1}$ 
are the vertical integration constants.

And finally,

(iv) The entropy generation equation:
$$
\Sigma v T \frac{ds}{dx} = Q^{+} - Q^{-},
\eqno(6)
$$
where, $T$, and $s$ are the temperature and entropy density of the 
accretion flow, respectively. $Q^{+}$ and $Q^{-}$ are the heat gain and lost by the 
flow, respectively. In the subsequent sections, we set $Q^{-} = 0$ as 
we ignore the cooling effect in this work. Following the prescription described in 
Aktar et al. 2017, we solve equations (1-6) in order to obtain the global transonic
accretion solutions around black holes.

\subsection{Equations for Outflow}

Already, we mentioned that the accretion flow geometry is confined in the equatorial 
plane and the jet or outflow geometry is described in the off-equatorial plane about 
the axis of rotation of black hole. As the jets  are tenuous in nature, the differential 
rotation of the outflowing matter is negligibly small. Thus, we ignore the viscous 
dissipation in the outflow calculation. Also, we consider the outflowing matter obey 
the polytropic equation of state as $P_j=K_j \rho_j^\gamma$, where, the suffix `$j$' 
denotes the outflow variables and $K_j$ is the measure of specific entropy of the jet. 
The equations of motion of the outflowing matter are given by,

(i) The energy conservation equation of outflow:
$$
\mathcal{E}_{j}=\frac{1}{2}v_j^2+\frac{a_j^2}{\gamma -1 }+\Phi_{eff},
\eqno(7)
$$
where, $\mathcal{E}_{j}$ ($\equiv {\mathcal E}$) represent the specific 
energy of the outflow, $v_j$ is the outflow velocity and $a_j$ is the sound 
speed of the outflow, respectively.

(ii) Mass conservation equation of outflow:
$$
{\dot{M}_{out}}=\rho_{j}v_{j}\mathcal{A}_j,
\eqno(8)
$$
where, $\mathcal{A}_j$ represents the total area function of the jet. The jet area 
function $\mathcal{A}_j$ can be calculated by knowing the radius of two boundary 
surfaces, namely the centrifugal barrier (CB) and the funnel wall (FW). The 
centrifugal barrier (CB) is obtained by defining the pressure maxima surface as $(d
\Phi^{eff}/dx)_{r_{CB}}=0$ and the funnel wall (FW) is the pressure minimum 
surface which is defined by the null effective potential as $\Phi^{eff}\vert_{r_{FW}}
=0$. We use the jet area function considering the projection effect similar as Aktar et 
al. (2017).

\section{Solution Methodology}

In our model, the outflowing matter emerges out between the 
centrifugal barrier (CB) and funnel wall (FW) surfaces due to the excess 
thermal gradient force at the shock. Therefore, we focus only on the accretion 
solutions which satisfy Rankine-Hugoniot (hereafter RH) shock conditions. The local
specific energy of the accretion flow is calculated by integrating the equation (1) as, 
$\mathcal{E}(x) = v^2/2 + a^2/(\gamma -1) +\Phi_{eff}$. While doing this, we consider
the shock to be thin and non-dissipative. With this, we therefore have the 
shock conditions which are given by,

(i) the conservation of energy flux:
$$
\mathcal{E_{+}} = \mathcal{E_{-}},
\eqno(9a)
$$
The quantities with subscripts `$+$' and `$-$' sign indicate the value of the 
variables after and before the shock, respectively. 

(ii) the conservation of mass flux:
$$
\dot{M}_{+} = \dot{M}_{-} - \dot{M}_{out}=\dot{M}_{-}(1 - R_{\dot{m}}),
\eqno(9b)
$$
where, $\dot{M}_{-}$, and $\dot{M}_{+}$ are the pre-shock, post-shock mass 
accretion rate. $\dot{M}_{out}$ is the outflowing mass rate, where mass outflow 
rate is defined as $R_{\dot{m}} =\dot{M}_{out}/\dot{M}_{-}$. 

(iii) the conservation of momentum flux:
$$
W_{+} + \Sigma_{+} v_{+}^2 = W_{-} + \Sigma_{-} v_{-}^2.
\eqno(9c)
$$ 
Now, as the inflow and outflow variables are coupled by the shock 
conditions, we solve accretion and outflow equations simultaneously
(Aktar et al, 2017).
Since outflowing matter emerges from the post-shock region (PSC), we 
assume that the jet launches with the same density as the density of the 
post-shock matter i.e., $\rho_{j}=\rho_{+}$. The mass outflow rates is 
calculated in terms of inflow and outflow properties at the shock using 
equations (2) and (8) and is given by,
$$
R_{\dot{m}}=\frac{\dot{M}_{out}}{\dot{M}_{-}} 
=\frac{Rv_j(x_s)\mathcal{A}_j(x_s)}{4\pi \sqrt{\frac{1}
{\gamma}} x_s^{3/2}{\Phi_{r}^{\prime}}^{-1/2} a_+ v_-} ,
\eqno(10)
$$
where, $R=\Sigma_+/\Sigma_-$, is the compression ratio, $v_j(x_s)$ and 
$\mathcal{A}_j(x_s)$ are the jet velocity and jet area function at the shock, 
respectively. Here, we use iterative method to calculate $R_{\dot{m}}$ 
self-consistently similar as Aktar et al. (2017).


\section{Results}
In this section, we present the results based on our model calculations. More 
precisely, we discuss the effect of viscosity on mass loss around rotating black hole. 
In this paper, we ignore the cooling effect. We also calculate the maximum QPO 
frequency based on our model. The adiabatic index $\gamma$ lies in the range 4/3 $
\le$ $\gamma$ $\le$ 5/3 as far as theoretical limit is concerned. In this work, we use 
$\gamma =1.4$ throughout our calculation.

\subsection{Effect of viscosity on mass loss}
\begin{figure}[!t]
\includegraphics[width=0.95\columnwidth]{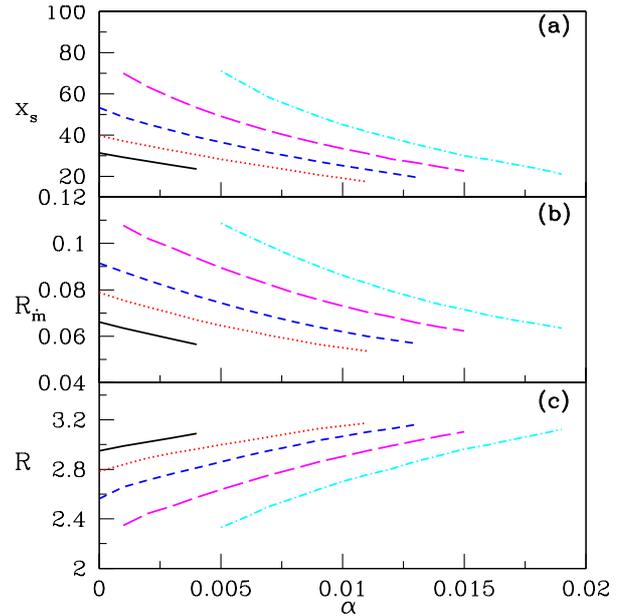}
\caption{Variation of $(a)$ shock location $(x_s)$, $(b)$ outflow rates $(R_{\dot{m}})$ and $(c)$ 
compression ratio $(R)$ as a function of viscosity parameter $\alpha$. Here, we vary spin 
$a_k$ = 0.36 to 0.52, with an interval $\Delta a_k=0.04$ from left most to right most curve.}
\label{fig One}
\end{figure}

In Figure \ref{fig One}, we present the effect of viscosity parameter on inflow-outflow solutions 
around rotating black holes. Here, we fix the injection radius at $x_{inj}=500$ and 
the corresponding flow energy  and flow angular momentum are $(\mathcal{E}_{inj}, 
\lambda_{inj})$= (0.002, 2.95) at the injection radius, respectively. We vary the spin 
$(a_k)$ of the black hole from 0.36 to 0.52 from left most to the right most curves 
with an interval of $\Delta a_k = 0.04$. We observe that the 
shock location $(x_s)$ moves towards the black hole horizon with the increase of 
viscosity parameter $\alpha$ in presence of mass loss, depicted in Figure 1a. Viscosity 
transports the angular momentum outward from the PSC, as a result, the centrifugal 
barrier weakens and this triggers the shock to move inward towards the black hole 
horizon. This also indicates that the effective area of PSC decreases with the increase 
of viscosity parameter $\alpha$. As a results, the amount 
of inflowing matter intercepted by the PSC is reduced and eventually, the outflow 
rate is also decreased with the increase of viscosity parameter. This is in 
agreement with the results reported by Chattopadhyay \& Das (2007). The variation of 
outflow rates $(R_{\dot{m}})$ is plotted in terms of viscosity parameter ($\alpha$) 
in Figure 1b corresponding to Figure 1a. Compression ratio indicates that how much the 
flow is compressed across the shock and it is measured by the ratio 
of post-shock density to the pre-shock density. The corresponding compression ratio 
$(R)$ is presented in Figure 1c. The compression ratio increases with the increase of 
viscosity because the shock forms closer to the black hole.

\begin{figure}[!t]
\includegraphics[width=.95\columnwidth]{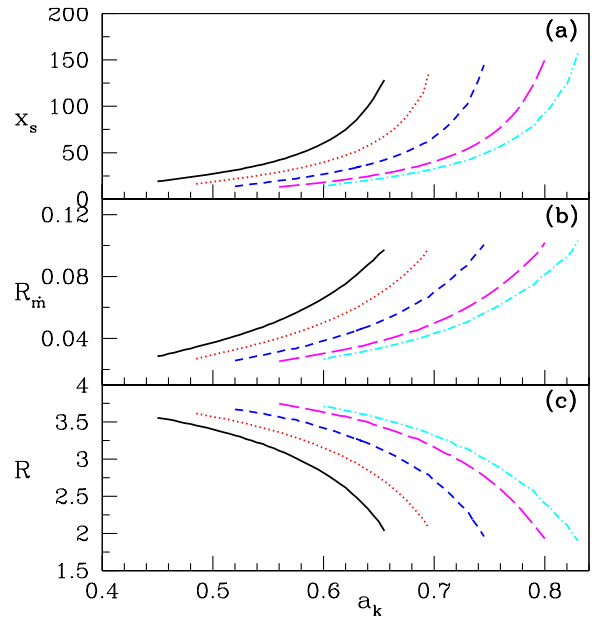}
\caption{Variation of $(a)$ shock location $(x_s)$, $(b)$ outflow rates $(R_{\dot{m}})$ and $(c)$ 
compression ratio $(R)$ as a function of spin of the black hole $a_k$. Here, we vary the viscosity 
parameter $\alpha$ = 0.001 to 0.009, with an interval $\Delta \alpha =0.002$ from left most to 
right most curve.}
\label{fig Two}
\end{figure}

In the same context, we show the effect of spin of the black hole on the shock 
properties and outflow rates in the viscous accretion flow in Figure \ref{fig Two}. Here, we fix the 
outer boundary at $x_{inj} = 500$ and flow energy and flow angular momentum $
(\mathcal{E}_{inj}, \lambda_{inj})$ = (0.0015, 2.85), respectively. We observe that 
the shock locations recede away from the black hole horizon with the increase of 
black hole spin $a_k$ for fixed viscosity parameter $\alpha$. As a consequence, the 
effective area of PSC also increases with the rotation of black hole. So, the amount of 
inflowing matter deflected by the PSC is also increased and ultimately, it increases 
the outflow rates with the spin of the black hole. But, for a fixed spin value, the 
outflow rates decreases with the increase of viscosity parameter similar to Figure 1b. 
On the other hand, the compression ratio decreases with the increase of spin of black 
hole. The variation of shock location $(x_s)$, outflow rates $(R_{\dot{m}})$ and 
compression ratio $(R)$ are shown in Figure 2(abc), 
respectively. All these findings are in agreement with the result of Aktar et 
al. (2015) in the context of inviscid accretion flow.

\begin{figure}[!t]
\includegraphics[width=.95\columnwidth]{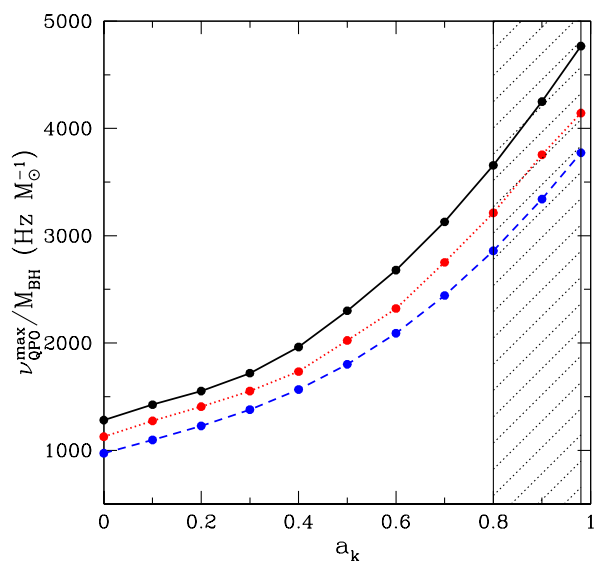}
\caption{Variation of maximum QPO frequency $(\nu_{QPO}^{max})$ for generalized black hole 
mass $(M_{BH})$ as a function of spin $a_k$. Solid, dotted and dashed curves represent $\alpha 
= 0.01$ (black), 0.05 (red) and 0.1 (blue), respectively. The shaded region represents the results 
beyond $a_k > 0.8$.}
\label{fig Three}
\end{figure}

\subsection{Estimation of QPO frequency}

Recent numerical simulation work of Das et al. (2014) reveals the fact that 
when viscosity parameter exceeds its critical limit, standing shock solution 
disappears and PSC exhibits quasi-periodic oscillation. In addition, Okuda \& Das
(2015) showed the unstable nature of the inner part of the advective
accretion disk for low angular momentum flow around black holes. Motivating
with this, in the present work, we calculate the QPO frequency considering the
accretion-ejection model of viscous advective accretion flow around rotating
black hole.

Here, we estimate the infall time from the post-shock velocity profile as $t_{infall}=\int{dt}
=\int_{x_s}^{x_{in}}\frac{dx}{v(x)}$, where $v(x)$ is the post-shock velocity. For this 
calculation, the integration is carried out from the shock location to the inner sonic 
point as the distance between the inner sonic point to the event horizon is negligibly 
small. Now, we estimate the QPO frequency from the infall time as $\nu_{QPO}\sim
1/t_{infall}$ in units of $r_g/c$ (Molteni et al. 1996b), where $r_g=GM_{BH}/c^2$. 
The unit of QPO frequency is converted into Hertz by multiplying with $c/r_g$. We vary all the flow parameters freely to calculate the 
QPO frequency scaled with black hole mass ($\nu_{QPO}/M_{BH}$~{\rm Hz} $M^{-1}_{\odot}$)
and then identify the maximum QPO frequency ($\nu_{QPO}^{max}/M_{BH}$~{\rm Hz} $M^{-1}_{\odot}$)
for a given $a_k$. Here $M_{BH}$ denotes  the black hole mass 
measured in units of solar mass ($M_{\odot}$). In this analysis, 
we notice that shocks corresponding to $\nu_{QPO}^{max}$ 
usually form very close to the black hole but not 
necessarily its closest position. In Figure 3, we plot $\nu_{QPO}^{max}/M_{BH}$ 
as function of $a_k$ where filled circles connected with solid, dotted and dashed 
curves are for viscosity $\alpha=0.01$ (black), $\alpha=0.05$ (red) and $\alpha=0.1$,
respectively. This maximum QPO frequency calculation is more general in the sense
that it is independent of black hole mass.

\section{HFQPOs and Spin of BH sources - Application to GRO J1655-40}

Several attempts have been made to constrain the spin (Shafee et al. 2006; Miller 
et al. 2009) using the characteristics
of the HFQPOs (Wagoner et al. 2001; Abramowicz et al. 2001) observed in Galactic 
black hole sources (Remillard et al., 1999; Strohmayer, 2001; Belloni et al., 2012). 
In our earlier paper, we attempted to constrain the spin of GRO J1655-40 using the 
observed maximum QPO frequency ($\sim$ 450 Hz). Based on our accretion-ejection 
formalism (see Aktar et al. 2017 for details), we further attempt to constrain 
the spin of GRO J1655-40 using both HFQPOs ($\sim$ 300 Hz and $\sim$ 450 Hz) observed 
in this source.

\subsection{Observation and Data Reduction}

The Galactic black hole source GRO J1655-40 has two characteristic HFQPOs observed
at $\sim$ 300 Hz and $\sim$ 450 Hz in the power spectrum. There have been several 
detections of the 300 Hz HFQPOs during the 1996 outburst of the source. The detection 
of 300 Hz HFQPOs (Remillard et al., 1999, Belloni et al., 2012) is not so significant, 
whereas the detection of 450 Hz HFQPO (Strohmayer, 2001, Belloni et al., 2012) is 
prominent and significant (see also Aktar et al. 2017). For our specific purpose, 
we re-analysed the archival RXTE data for both HFQPOs ($\sim$ 300 Hz and $\sim$ 450 Hz). 
We use two RXTE observations which was carried out on 01 August, 1996 
(ObsID: 10255-01-04-00) and on 09 September, 1996 (ObsID: 10255-01-10-00) for the
detection of HFQPOs at $\sim$ 300 Hz and $\sim$ 450 Hz respectively.
For both observations, light curves were generated with a time resolution of 
0.00048828125 sec with Nyquist frequency of 1024 Hz in the energy band of 2-12 keV, 
13-27 keV and 2-27 keV. 
We extract light curve from science array data (exposure of $\sim$ 9 ksec) for the observation on 
01 August, 1996 and from science event data (exposure of $\sim$ 8 ksec) for the observation 
on 09 September, 1996 based on required time resolution and energy ranges. 
Further, we follow the standard procedures (see Nandi et al. 2012; Radhika et al. 2016a) 
to extract the energy spectrum in the energy range of $3 - 150$ keV using both PCA ($3-30$ keV) 
and HEXTE ($25-150$ keV) data of RXTE observations.
 
\subsection{Analysis and Modelling}

\begin{figure}[!t]
\includegraphics[width=1.05\columnwidth]{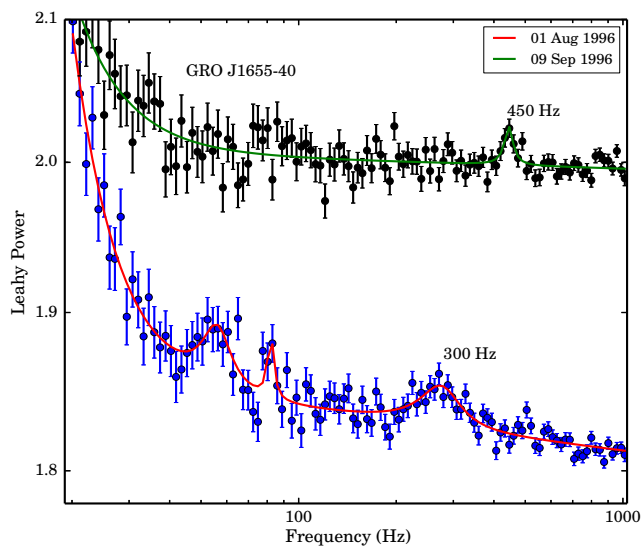}
\caption{Power spectra corresponding to the $\sim$ 300 Hz and $\sim$ 450 Hz HFQPOs of 
GRO J1655-40 during its 1996 outburst are shown in {\it Lehay Power} - {\it Frequency} 
space. The QPO like features are modelled with \textit{lorentzians}. Detection of 
HFQPOs at $\sim$ 300 Hz and $\sim$ 450 Hz are marked in the plot.}
\label{fig Four}
\end{figure}

We compute  the power spectra for both observations in the range 20 Hz to 
1024 Hz with 8192 newbins per frame and with a geometrical binning factor of -1.02. 
The 300 Hz QPO was detected only in the energy range 2-12 keV while the 450 Hz HFQPO 
was observed in the energy range 13-27 keV. The power spectra are 
modeled with a \textit{lorentzian} and \textit{powerlaw} components for QPO like 
feature and continuum in the lower frequency range respectively. Figure \ref{fig Four} 
shows the simultaneous plot of both HFQPOs observed in GRO J1655-40 during 1996 outburst.

The $\sim 300$ Hz HFQPO has centroid frequency at $275\pm 9$ Hz (see Figure 4) 
with an FWHM of 71.7 Hz. The detection of broad feature has a significance 
of $\sim$ 4.17 and a Q-factor of $\sim$ 3.84. Power spectrum also shows two more 
QPO like feature at around 57 Hz and 80 Hz.
The highest observed HFQPO (see Figure 4) was found at a central frequency 
of 444.8 Hz with an FWHM of 32.4 Hz, a significance of 2.79 and a Q-factor of 13.72. 
Both HFQPOs plotted in Figure 4 is in {\it Leahy Power} - {\it Frequency} space 
(Leahy et al. 1983).

\begin{figure}[!t]
\includegraphics[angle = 270, width=0.48\textwidth]{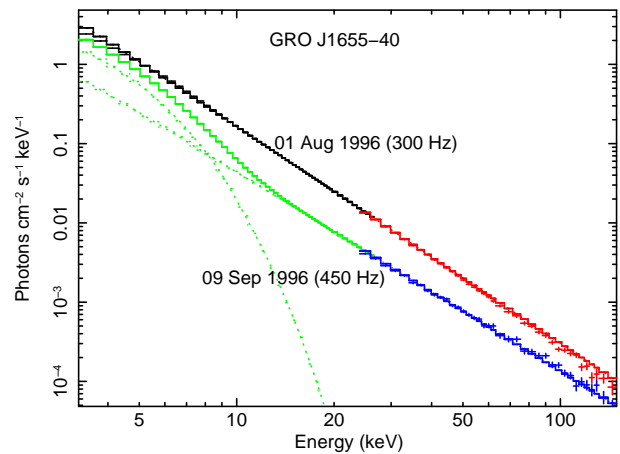}
\caption{Broadband energy spectra (3-150 keV) using PCA and HEXTE data 
corresponding to the $\sim$ 300 Hz and $\sim$ 450 Hz HFQPOs are observed during 
1996 outburst of GRO J1665-40. Both energy spectra show strong high energy emission
with steep powerlaw component. Interestingly, the energy spectrum of 300 Hz observation
shows very weak signature of thermal emission, whereas thermal emission 
is strong in the energy spectrum of 450 Hz observation. See text for details.}
\label{fig Five}
\end{figure}

In Figure \ref{fig Five}, we plot the combined broadband energy spectra of 
both observations in the energy range of 3 - 150 keV.
We modelled the energy spectra with phenomenological models consisting of 
\textit{diskbb} and a \textit{powerlaw} components along with \textit{phabs} model
to consider the equivalent hydrogen column density ($nH \sim 0.7 \times 10^{22}$
atoms ${\rm cm}^{-2}$) in the line-of-sight towards the source. 

The energy spectrum of $\sim$ 300 Hz observation (01 August, 1996) was modelled 
with a steep \textit{powerlaw} component (photon index, $\Gamma=$ 2.80) in the energy
range of $3-150$ keV. A very weak signature of thermal emission is observed at 
low energies. 
Additionally, a smeared edge at $\sim$ 8 keV was required for fitting the spectrum. 
On the other hand, the energy spectrum of 450 Hz QPO (09 September, 1996) was modelled 
with a strong disk component (see Figure 5) of disk temperature of 
$\sim$ 1.3 keV and with considerably steeper powerlaw ($\Gamma=$ 2.43) at higher
energy. This also clearly indicates that the disk has proceeded towards the black hole horizon with a smaller size of PSC for 450 Hz QPO compared to the 300 Hz QPO.

\begin{figure}[!t]
\includegraphics[width=.95\columnwidth]{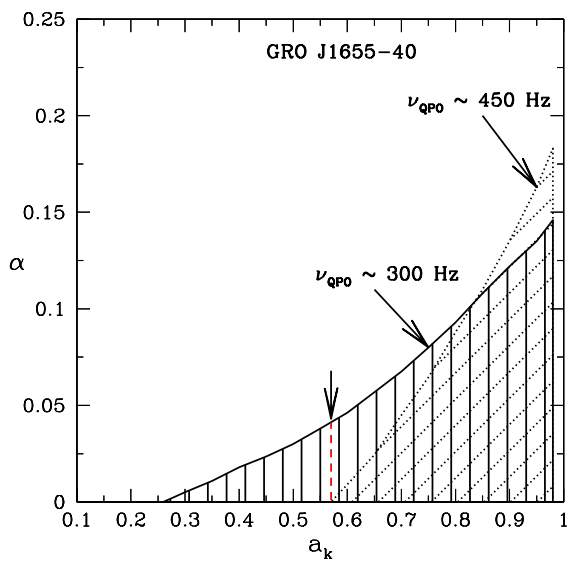}
\caption{Ranges of viscosity parameter $(\alpha)$ and black hole spin $(a_k)$ that 
provide HFQPOs $\nu_{QPO}$ $\sim$ 300 Hz and $\sim$ 450 Hz observed in GRO J1655-40.}\label{fig Six}
\end{figure}

Following the prescription of Aktar et al. 2017, we identify the parameter
spaces spanned by the black hole spin ($a_k$) and viscosity ($\alpha$) that
successfully reproduce the HFQPOs ($\sim$ 300 Hz and $\sim$ 450 Hz) 
observed for the black hole source GRO J1655-40. While doing so,
we vary all the remaining flow variables freely and  consider the mass 
of GRO J1655-40 as $6 M_{\odot} $ (Green et al. 2001; Beer \& Podsiadlowski 2002).
In Figure 6, the region shaded with solid and dotted lines denote the parameter space
corresponding to $\sim 300$ Hz and $\sim 450$ Hz. The latter result is recently 
reported by Aktar et al, 2017, where authors constrained the spin of GRO J1655-40
and found the minimum value $a_k^{\rm min} \sim  0.57$ as indicated by the dashed vertical 
(red) line with downward arrow at the top. The viscosity associated with $a_k^{\rm min}$
is obtained as $\alpha \sim 0$.
Since black hole spin is not expected to evolve noticeably within the 
couple of weeks or so, it is reasonable to adhere the above spin range and calculate the 
range of viscosity as function of $a_k$ that provides $\sim 300$ Hz QPOs. 
We find that viscosity spreads over a wide range for all $a_k$ and in particular, 
we obtain $0 \le \alpha <0.04$ for $a_k^{\rm min} \sim  0.57$. 
When the apriori restriction on $a_k$ is removed, we find the spin ranges 
for $\sim 300$ Hz QPO is $a_k$ $\ge 0.26$ which is in agreement with 
the results reported by Abramowicz \& Kluz\'niak 2001; Motta et al. 2014; Stuchl\`ik \& 
Kolo\`s 2016. Since there exists a common overlapping region in  the $a_k - \alpha$
plane, we argue that the spin of the black hole source GRO J1655-40 seems to be $a_k \ge 0.57$.

\section{Conclusions}

We self-consistently solve the accretion-ejection hydrodynamic equations around
rotating black holes by using the inflow parameters and subsequently calculate the 
mass outflow rates. We observe that viscosity plays 
an important role in deciding the mass loss from the advective accretion disk. As the
viscous dissipation is increased, shock location (equivalently size of PSC) as well as 
mass outflow rate are decreased for a given spin value. Since PSC is hot and dense, 
hard radiations are likely to emerge out from PSC. This essentially
indicates that the dynamics of PSC manifests both the mass outflow rates and
spectro-temporal properties of the black hole.  

Next, we explore the 
characteristics of HFQPOs ($\sim$ 300 Hz and $\sim$ 450 Hz)
and corresponding broadband energy spectra of the source GRO J1655-40. We 
find that the low energy photons ($2-12$ keV) contribute to generate the 
$\sim 300$ Hz QPO, whereas $\sim 450$ Hz QPO has been observed in the higher energy 
band of $13-27$ keV. Moreover, the energy spectrum of $\sim 300$ Hz QPO does not
show any prominent signature of disk emission, although the disk contribution  
(thermal emission) is significant in the spectrum (upto $15$ keV) containing $\sim 450$ Hz QPO.
It has also been observed that both the energy spectra are dominated by steep 
{\it powerlaw} components ($\Gamma= 2.80$ and $2.43$ for $\sim 300$ Hz and $\sim 450$ Hz, 
respectively) up to the high energies ($\sim 150$ keV).

The spectro-temporal analysis indicates that the origin of HFQPOs in GRO J1655-40
probably may not be coupled with disk emission ($i. e.$, independent of disk flux).
Rather, it is possible that a strong source might be present in the system, which
produces not only the high energy photons, but also responsible for HFQPOs 
observed in this source. Based on the above discussion, it is quite indicative
that PSC behaves like the strong source that inverse-comptonizes the significant
fraction of the soft photons originated from the disk and produces the high energy 
emission with steep power-law index.
Moreover, when PSC forms close to black hole and exhibits modulation, it 
gives rise to HFQPOs which we observe for GRO J1655-40. 
This apparently provides an indication that the 
origin of HFQPOs in black hole systems could be explained using the method of
shock dynamics without invoking the prescription of orbital frequency 
associated with inner most stable circular orbit (Strohmayer, 2001).
Further, we discuss the limiting range of the spin parameter of GRO J1655-40
based on the observed HFQPOs.

\section*{Acknowledgement}
Authors thank the reviewer for useful comments and suggestions.
AN thanks GD, SAG; DD, PDMSA and Director, ISAC for encouragement and continuous
support to carry out this research.
This research has made use of the data obtained through High Energy Astrophysics
Science Archive Research Center online service, provided by NASA/Goddard Space
Flight Center.



\end{document}